\documentclass[aps,twocolumn,floatfix,superscriptaddress]{revtex4}
\usepackage{epsfig}

\begin{document}

\title{Self-affine surface morphology of plastically deformed metals}

\author{Michael Zaiser}
\affiliation{Center for Materials Science and Engineering, The
University of Edinburgh, The King's Buildings, Sanderson Building,
Edinburgh EH93JL, UK}
\author{Frederic Madani}
\affiliation{Center for Materials Science and Engineering, The
University of Edinburgh, The King's Buildings, Sanderson Building,
Edinburgh EH93JL, UK} \affiliation{Laboratory of Mechanics of
Materials, Aristotle University of Thessaloniki, 54006
Thessaloniki, Greece}
\author{Vasileios Koutsos}
\affiliation{Center for Materials Science and Engineering, The
University of Edinburgh, The King's Buildings, Sanderson Building,
Edinburgh EH93JL, UK}
\author{Elias C. Aifantis}
\affiliation{Laboratory of Mechanics of Materials, Aristotle
University of Thessaloniki, 54006 Thessaloniki, Greece}

\begin{abstract}
We analyze the surface morphology of metals after plastic
deformation over a range of scales from 10 nm to 2 mm, using a
combination of atomic force microscopy and scanning white-light
interferometry. We demonstrate that an initially smooth surface
during deformation develops self-affine roughness over almost four
orders of magnitude in scale. The Hurst exponent $H$ of
one-dimensional surface profiles is initially found to decrease
with increasing strain and then stabilizes at $H \approx 0.75$. By
analyzing their statistical properties we show that the
one-dimensional surface profiles can be mathematically modelled as
graphs of a fractional Brownian motion. Our findings can be
understood in terms of a fractal distribution of plastic strain
within the deformed samples.
\end{abstract}

\pacs{62.20.Fe,68.35.-p,89.75.-k} \maketitle

According to the traditional paradigm implicit in continuum models
of plasticity, plastic deformation under homogeneous loads and in
the absence of so-called plastic instabilities is expected to
proceed in a smooth and spatially homogeneous manner.
Fluctuations, if any, are supposed to average out above the scale
of a ``representative volume element'' which is tacitly assumed to
be small in comparison with macroscopic dimensions (or, indeed, in
comparison with the scale of any problem to which continuum
modelling is applied). Recently for crystalline systems deforming
by dislocation glide this paradigm has been challenged both from
the theoretical and experimental side
\cite{MIG01A,MIG01B,WEI01,ZAI01,MIG02,WEI03}. Weiss et. al.
demonstrated that dislocation glide in single-glide oriented ice
single crystals proceeds in a temporally intermittent manner,
i.e., plastic flow is characterized by a sequence of large
``deformation bursts'' \cite{MIG01A,WEI01}. Acoustic emission due
to these bursts was monitored and it was demonstrated that there
is no characteristic burst magnitude. Instead the energy releases
were found to obey a scale-free power-law distribution over more
than six orders of magnitude in energy \cite{MIG01A}. Simulations
of dislocation dynamics in single-glide single crystals showed
qualitatively similar behavior \cite{MIG01A,MIG01B}. These
findings were interpreted as dynamical critical phenomena in a
slowly driven non-equilibrium system \cite{ZAI01,MIG02}, and it
has been argued that the dislocation system in a plastically
deforming crystal is always close to a critical point (``yielding
transition'', \cite{ZAI01}).

Temporal intermittency of plastic flow goes along with spatial
heterogeneity. This has been long known to metallurgists who
observed pronounced surface traces (``slip lines'') resulting from
the collective motion of many dislocations. By monitoring the
evolution of surface features it is therefore possible to study
collective behavior in plastic flow. However, most investigations
(for an overview, see \cite{NEU83}) have focused on the evolution
of single slip lines or slip line bundles, i.e., the motion of
isolated dislocation groups. Long-range correlations within the
slip pattern have only occasionally been studied. Konstantinidis
and Aifantis \cite{KON02} used a wavelet transform to investigate
the slip pattern of a Fe$_3$Al single crystal on multiple scales
between 0.1 and 50 micrometers but did not quantitatively analyze
their data in view of scale-invariant behavior. Kleiser and Bocek
\cite{KLE86} report a self-similar pattern of slip lines in Cu
single crystals on scales between 0.06 and 2 micrometers and
determine a fractal dimension $D_{\rm F} \approx 0.5$ for the set
of intersection points between the slip lines and a line normal to
the slip direction.

Recently, Weiss and Marsan \cite{WEI03} investigated the
three-dimensional patterning of slip on macroscopic scales. By
recording acoustic emission with multiple transducers and using
spatial triangulation, they determined the 3D distribution of slip
bursts in the volume of an ice single crystal. This investigation
demonstrated that the slip burst pattern on macroscopic scales
(0.6 mm to 10 mm) exhibits features of a fractal set with
correlation dimension $D_{\rm C} \approx 2.5$. However, due to the
limited spatial resolution of the technique no information about
the slip distribution on smaller scales could be obtained.

In the present study we pursue a slightly different approach for
assessing spatial correlations in plastic deformation from the
nanoscopic to the macroscopic scale. Instead of directly studying
the distribution of plastic strain in the bulk or on the surface
of the sample, we monitor the evolution of the surface profile.
Provided that the initial surface is sufficiently smooth, the
evolution of a one-dimensional profile can be directly related to
the plastic distortion at the surface: If we define a local
coordinate system such that the $x$ direction corresponds to the
direction of the profile and the $y$ direction to the surface
normal, then the derivative $y_x = \partial y/\partial x$ of the
profile $y(x)$ equals the component $\beta_{yx}$ of the plastic
distortion tensor. Long-range correlations in the plastic strain
pattern can be detected since they give rise to characteristic
modifications of the surface morphology - more specifically
speaking, power-law correlations in the spatial distribution of
plastic strain lead to the emergence of self-affine surface
profiles with a Hurst exponent $H > 0.5$ as will be discussed
below.

A self-affine profile $y(x)$ is characterized by the property that
the statistical properties of the profile remain unchanged under
the scaling transformation $x \to \lambda x, y \to \lambda^H y$.
The roughness exponent or Hurst exponent $H$ can be related to a
fractal dimension $D_{\rm F}$ (``box dimension'') of the profile
through $D_{\rm F} = 2 - H$ \cite{MAN85}. Self-affine behavior is
abundant in natural surfaces, a most prominent example in
materials physics are fracture surfaces whose large-scale
roughness can be characterized by a universal
(material-independent) exponent $H \approx 0.8$ \cite{BOU97}. In
the present letter we report for the first time an analysis of the
surface structure of plastically deformed metal samples in terms
of self-affine properties. We applied a combination of atomic
force microscopy (AFM) and scanning white-light interferometry
(SWLI) to quantitatively characterize the surface morphology over
a range of scales between 10 nm and 2 mm. Polycrystalline copper
samples of 99.9 \% nominal purity and an average grain size of 40
$\mu$m were electropolished to a typical rms roughness of 44 nm as
determined by SWLI over a surface area of $40 \mu$m $\times 40
\mu$m. Two complementary methods were used for surface imaging:
(i) AFM scans (512 $\times$ 512 pixels) were taken over areas of
$26 \mu$m $\times 26\mu$m and $6 \mu$m $\times 6 \mu$m using a
PicoSPM (Molecular Imaging) at the constant deflection mode with a
medium range scanner and a cantilever (Mikromasch, Ultrasharp)
with a nominal spring constant of 1.75 N/m and tip curvature
radius ~10 nm; the rms vertical noise amplitude was less than 0.05
nm; (ii) SWLI (New View 100, Zygo) scans were performed over areas
of $130 \mu$m $\times 2000 \mu$m with a lateral resolution of ~0.6
$\mu$m and a vertical resolution of ~0.5 nm. From the scans,
typically 4-5 surface profiles were taken in the direction of the
specimen axis and 2-3 profiles normal to that direction.

\begin{figure}[t]
\centerline{\epsfig{file=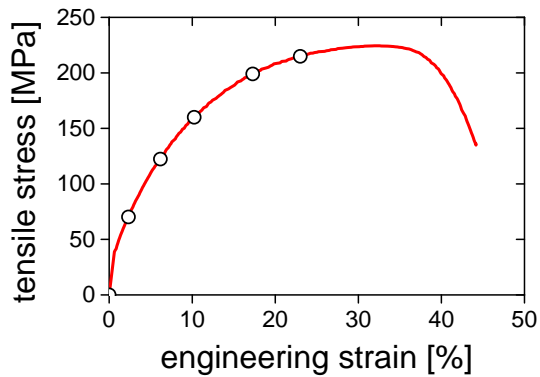,width=7cm,clip=!}}
\caption{Stress-strain curve of a polycrystalline Cu specimen;
deformation at room temperature, imposed deformation rate during
loading $\dot{\epsilon} = 1.5 \times 10^{-3}$ s$^{-1}$; surface
profiles were taken before deformation as well as after unloading
at the points indicated by the open circles.} \label{stressstrain}
\end{figure}
After characterizing the initial surface, the samples were
deformed in tension on a standard tensile-testing machine (Instron
series 3360). Testing was done at room temperature with an imposed
strain rate $\dot{\epsilon} = 1.5 \times 10^{-3}$ s$^{-1}$. At
total (engineering) strains $\epsilon = 2.3\%, 5.5 \%, 9.6 \%,
17.8 \%$ and $23 \%$ (see Figure \ref{stressstrain}), the sample
was unloaded and AFM and SWLI scans were taken in a similar manner
as before deformation. The sequence of surface morphology
investigations was ended at the onset of macroscopic deformation
localization (necking).

\begin{figure}[t]
\epsfig{file=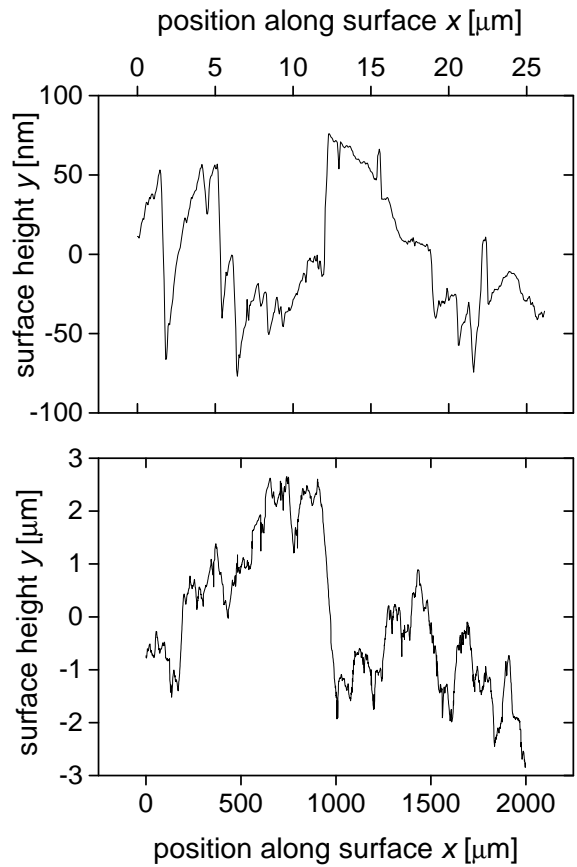,width=7.5cm,clip=!} \caption{Surface
profiles taken at $\epsilon = 9.6 \%$; top: AFM profile; bottom:
SWLI profile; the $x$ direction is parallel to the direction of
the tensile axis; roughness plots corresponding to these profiles
are shown in Figure \ref{roughness}.} \label{profiles}
\end{figure}
\begin{figure}[tbh]
\centerline{\epsfig{file=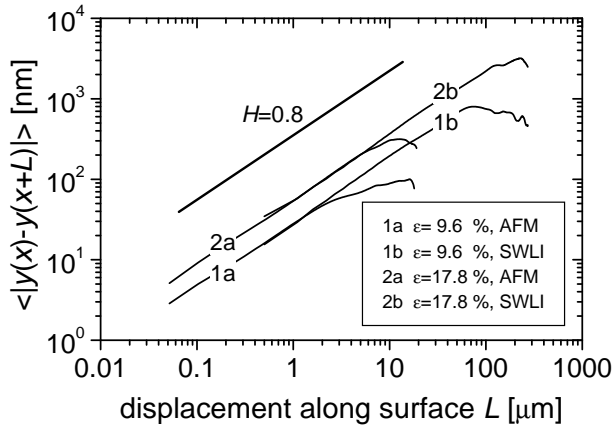,width=8cm,clip=!}}
\caption{Roughness plots (mean height difference vs. distance
along profile) for AFM and SWLI profiles obtained at strains of
9.6 and 17.8 \%; the corresponding profiles for $\epsilon = 9.6
\%$ are shown in Figure \ref{profiles}.}\label{roughness}
\end{figure}

Typical surface profiles are shown in Figure \ref{profiles}
representing a ``long'' AFM profile obtained from a $26 \mu$m
$\times 26 \mu$m scan together with a SWLI profile. To investigate
a possible self-affine structure, mean height differences $\langle
|y(x)-y(x+L)| \rangle$ were evaluated  as a function of $L$ for
each of the AFM and SWLI profiles by averaging over all pairs
$(x,x+L)$ within the profile. Self-affine behavior implies that
\begin{equation}
\langle |y(x)-y(x+L)|\rangle \propto L^H\;.
\end{equation}
Hence, double-logarithmic plots of $\langle |y(x)-y(x+L)| \rangle$
vs. $L$  should exhibit a linear scaling regime with slope $H$.
Such 'roughness plots' are shown in Figure \ref{roughness}.
Typical plots exhibit linear scaling regimes extending between
0.05 and 5 $\mu$m for the AFM and between 0.5 and 100 $\mu$m for
the SWLI profiles. Hurst exponents deduced from the slope of the
scaling regimes are similar for AFM and SWLI profiles, and also
the absolute values of $\langle |y(x)-y(x+L)| \rangle$ are similar
across the region of overlap of the roughness plots. Hence, we
observe continuous scaling over almost four orders of magnitude.
Surface profiles $y(z)$ taken in the direction normal to the
tensile axis exhibit scaling with similar exponents, although the
length of the self-affine scaling regimes is reduced.
Roughness plots obtained from the ``short'' AFM profiles deduced
from $6\mu$m $\times 6 \mu$m scans show a crossover to a
strain-independent apparent Hurst exponent $H \approx 1$ below
50-100 nm.

\begin{figure}[t]
\centerline{\epsfig{file=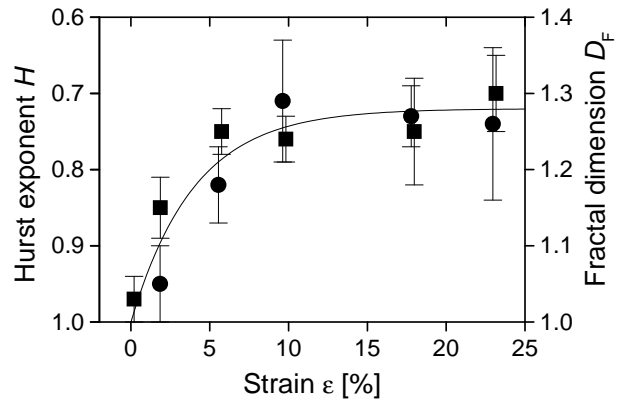,width=8cm,clip=!}}
\caption{Fractal dimension and Hurst exponent as a function of
strain, squares: values determined from scanning interferometry
profiles, circles: values determined from AFM profiles; each value
represents an average over 5 profiles; full line: guide to the
eye.}\label{dimension}
\end{figure}

Hurst exponents $H$ and fractal dimensions $D_{\rm F} = 2 - H$
determined from the roughness plots are compiled in Figure
\ref{dimension}. The error bars reflect the scatter of exponents
obtained from different profiles taken at the same strain. It can
be seen that the exponents initially decrease with increasing
strain and then stabilize above $\epsilon \approx 10 \%$ at a
value of $H \approx 0.75$.
It is important to note that the initial value of $H \approx 0.97$
does not reflect any self-affine surface roughness but simply
stems from large-scale gradients on an almost smooth surface. The
same is true for the steepening of roughness plots obtained from
the ``short'' AFM profiles on very small scales: Below 50 nm, the
apparent Hurst exponent increases towards $H \approx 1$ as one
essentially observes the smooth initial surface in between the
deformation-induced surface steps.

To further investigate the statistical surface properties in the
different scale regimes, probability distributions $p(\Delta y_L)$
of surface height differences $\Delta y_L = y(x) - y(x+L)$ have
been determined for the profiles taken at $\epsilon = 9.6 \%$.
$L_0 = 0.5 \mu$m was taken as a reference and all height
differences $\Delta y_L$ were normalized by the variance $y_0 =
\langle \Delta y_{L_0}^2 \rangle^{0.5} = 27.08$ nm of the
corresponding height difference distribution. Within the scaling
regime $\xi_0 < L < \xi_1$ ($\xi_0 \approx 50$ nm, $\xi_1 \approx
100 \mu$m), probability distributions corresponding to
\begin{figure}[tb]
\centerline{\epsfig{file=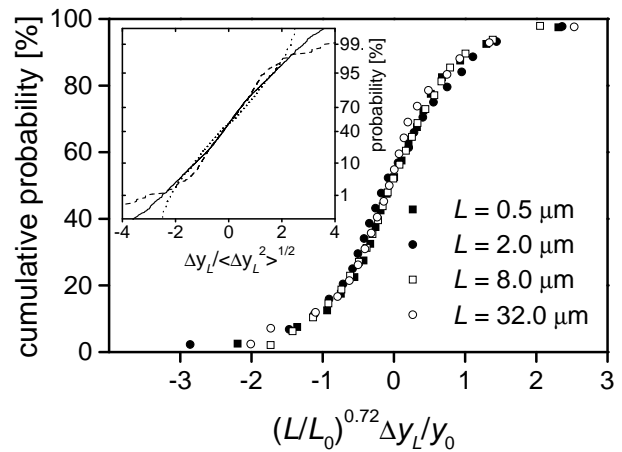,width=8cm,clip=!}}
\caption{Collapse of probability distributions of height
differences after re-scaling; data points: distributions for
different values of $L$ obtained from AFM and SWLI profiles at
$\epsilon = 9.6 \%$; insert: distributions (normalized to unit
variance) for $L = 40$nm (dashed line), $L =5 \mu$m (full line)
and $L= 160 \mu$m (dotted line).}\label{stat}
\end{figure}
different values of $L$ can be collapsed by re-scaling $\Delta
y_L/y_0 \to (L/L_0)^H \Delta y_L/y_0$ with $H = 0.72$ (see Figure
\ref{stat}):
\begin{equation}
p([L/L_0]^H \Delta y_L/y_0) = p(\Delta y_{L_0}/y_0)\;.
\end{equation}
To assess the shape of the probability distributions, cumulative
distributions normalized to unit variance have been represented on
a probability scale (i.e., the error function appears as a
straight line). The insert of Figure \ref{stat} shows the
cumulative distribution for $L=2 \mu$m (full line) as well as two
distributions for $L$ values outwith the scaling regime, namely
for $L = 40$nm (dashed line) and $L = 160 \mu$m (dotted line).
Whereas distributions throughout the self-affine scaling regime
are well approximated by Gaussian statistics, the tails of
distributions outwith the scaling regime exhibit characteristic
deviations from Gaussian behavior. On scales below $\xi_0$ the
surface morphology is characterized by flat regions separated by
huge slip steps at locations where slip events have occurred. This
leads to a leptokurtic distribution with heavy tails. On scales
above $\xi_1$, on the other hand, the surface flattens and large
height fluctuations are suppressed since the macroscopic slope of
the SWLI profiles is by definition set to zero. This gives rise to
platykurtic distributions with truncated tails.

The fact that the statistics is Gaussian throughout the scaling
regime implies that the surface profiles $y(x)$ can be
mathematically interpreted as graphs of fractional Brownian
motions. This allows us to draw conclusions on the underlying
plastic strain pattern since the increments $y_x$ of a fractional
Brownian motion exhibit long-range correlations given by
\cite{ENR04}
\begin{equation} \langle y_x(x) y_x(x')
\rangle \propto |x-x'|^{2H - 2}\;.
\end{equation}
Since $y_x$ equals the component $\beta_{yx}$ of the plastic
distortion tensor at the surface, our observations yield strong
evidence for power-law correlations in the distribution of plastic
strain. From the correlation integral
\begin{equation}
I(R) = \int_{|x-x'| < R} \langle y_x(x) y_x(x') \rangle {\rm
d}(x-x') \propto R^{2H-1}
\end{equation}
we deduce a correlation dimension $D_{\rm C} = 2H - 1 \approx 0.5$
for the one-dimensional strain pattern underlying the surface
profile. The value $D_C \approx 0.5$ is consistent with the
fractal dimension of slip line patterns reported by Kleiser and
Bocek \cite{KLE86}. A dimension of 0.5 for a one-dimensional
strain pattern is also consistent with the dimension $D_F = 2.5$
for the three-dimensional pattern of slip events measured by Weiss
and Marsan \cite{WEI03}. These investigations were performed on
single crystals; due to experimental limitations the scaling
regimes cover only about one order of magnitude. Our results
demonstrate that the strain pattern in polycrystals exhibits an
analogous self-similar behavior which by combining different
experimental techniques can be traced over almost four orders of
magnitude in scale.

More investigations are required to decide whether the Hurst
exponent $H \approx 0.75$ which we determine for deformed metal
surfaces is universal, i.e. independent of material and
deformation conditions. Further investigations are also needed to
establish the nature of the boundaries of the self-affine scaling
regime. The lower bound of the scaling regime can be related to
the spacing between individual 'slip events' between which the
surface remains unchanged by the deformation. The upper boundary
$\xi_1$ of the scaling regime, which defines the correlation
length of the self-affine surface pattern, is of the same order of
magnitude as the grain size (40 microns). One may reasonably
conjecture that the range of correlations in the dislocation
dynamics, and therefore in the slip pattern, is limited by the
grain size. However, further experimental work (preferably on
single crystals) is required to decide whether this is indeed the
physical mechanism which determines the correlation length.

From the theoretical side, the fact that the Hurst exponent of
deformed metal surfaces is quite similar to that of fracture
surfaces raises the fundamental question whether there is a
universal mechanism underlying both observations, or whether the
similarity of exponents is incidental. Crystal plasticity is a
bulk phenomenon which only indirectly affects the surface, whereas
fracture is directly governed by surface and interface properties.
However, this does not preclude the possibility of a common
theoretical treatment within the framework of non-equilibrium
critical phenomena.

In view of plasticity models using coarse-grained strain
variables, our results indicate that strain fluctuations may
average out rather slowly. The strain fluctuation on scale $L$ can
be estimated as $\Delta \epsilon_L \sim \Delta y_L/L$. At the
lower end of the self-affine scaling regime ($\xi_0 \approx 50$
nm), fluctuations are of the same order of magnitude as the
average strain. Within the scaling regime, strain fluctuations
decrease like $(\xi_0/L)^{1-H}$ if averaged over the length $L$;
for $(1-H) \approx 0.25$ we find that the residual strain
fluctuations at the upper end of the scaling regime ($\xi_1
\approx 100 \mu$m) are still about 15 \% of the average strain.
This observation explains why it is possible to detect significant
strain fluctuations even on macroscopic scales \cite{DIE56,ZAI96}.

\section*{Acknowledgments}

Financial support by the European Commission under RTN/DEFINO
HPRN-CT 2002-00198 and of EPSRC under Grant Nos. GR/S20406/01 and
GR/R43181/01 is gratefully acknowledged.


\begin{thebibliography}{47}

\bibitem{MIG01A}
M.-C. Miguel, A. Vespignani, S. Zapperi, J. Weiss and J. R.
Grasso, Nature {\bf 410}, 667 (2001).

\bibitem{MIG01B}
M.-C. Miguel, A. Vespignani, S. Zapperi, J. Weiss and J. R.
Grasso, Mater. Sci. Engng. A {\bf 309-310}, 324 (2001).

\bibitem{WEI01}
J. Weiss, J. R. Grasso, M.-C. Miguel, A. Vespignani and S.
Zapperi, Mater. Sci. Engng. A {\bf 309-310}, 360 (2001).

\bibitem{ZAI01}
M. Zaiser, Mater. Sci. Engng. A {\bf 309-310}, 304 (2001).

\bibitem{MIG02}
M.-C. Miguel, A.Vespignani, M.Zaiser and S. Zapperi, Phys. Rev.
Letters {\bf 89}, 165501 (2002).

\bibitem{WEI03}
J.Weiss and D.Marsan, Science {\bf 299}, 89 (2003).

\bibitem{NEU83}
H. Neuhauser, Slip-line formation and collective dislocation
motion, in {\em Dislocations in Solids}, edited by F.R.N. Nabarro
(North-Holland, Amsterdam, 1983), p. 319.

\bibitem{KON02} A.A. Konstantinidis and E.C. Aifantis, J. Engng.
Mater. Technol. {\bf 124}, 358 (2002).

\bibitem{KLE86} T. Kleiser and M. Bocek, Z. Metallkde. {\bf 77},
582 (1986).

\bibitem{MAN85} B. Mandelbrot, Phys. Scripta {\bf 32}, 257 (1985).


\bibitem{BOU97} E. Bouchaud, J. Phys: Cond. Mat. {\bf 9}, 4319
(1997).

\bibitem{ENR04} N. Enriquez, Stochast. Proc. Appl. {\bf 109}, 203
(2004).

\bibitem{DIE56} J. Diehl, Z. Metallkde. {\bf 47}, 331; 411 (1956).

\bibitem{ZAI96} M. Zaiser and P. H\"ahner, Phil. Mag. Letters {\bf
73}, 369 (1996).

\end{thebibliography}
\end{document}